# Doping effects on trimerization and magnetoelectric coupling of single crystal multiferroic (Y,Lu)MnO$_3$

Seongil Choi[1,2], Hasung Sim[1,3], Sunmin Kang[1,3], Ki-Young Choi[1], and Je-Geun Park[1,3*]

1. Center for Correlated Electron Systems, Institute for Basic Science (IBS) Seoul 08826, Korea
2. Department of Physics, Sungkyunkwan University, Suwon 16419, Korea
3. Department of Physics & Astronomy, Seoul National University, Seoul 08826, Korea

* Corresponding author: jgpark10@snu.ac.kr

### Abstract

Hexagonal RMnO$_3$ is a multiferroic compound with a giant spin-lattice coupling at an antiferromagnetic transition temperature [1]. Despite extensive studies over the past two decades, however, the origin and underlying microscopic mechanism of the strong spin-lattice coupling remain still very much elusive. In this study, we have tried to address this problem by measuring the thermal expansion and dielectric constant of doped single crystals Y$_{1-x}$Lu$_x$MnO$_3$ with x = 0, 0.25, 0.5, 0.75, and 1.0. From these measurements, we confirm that there is a progressive change in the physical properties with doping. At the same time, all our samples exhibit clear anomalies at T$_N$, even in the samples with x = 0.5 and 0.75 as opposed to some earlier ideas, which suggests an unusual doping dependence of the anomaly. Our work reveals yet another interesting facet of the spin lattice coupling issue in hexagonal RMnO$_3$.

## 1. Introduction

Multiferroics refers to multifunctional materials that possess simultaneously two or more ferroic orders like ferromagnetism, ferroelectricity and/or ferroelasticity [1-8]. However the coexistence of more than two orders is not common in materials, and the coupling among them is usually very weak even if there is such a coexistence of two orders in a single compound [9-11]. Of interesting multiferroic materials, rare-earth hexagonal manganites $R$MnO$_3$ ($R$ = rare-earth ions) is known to have a ferroelectric transition temperature above 900 K and an antiferromagnetic transition temperature around 100 K [12, 13].

The crystallographic structure of $R$MnO$_3$ with a relatively smaller ionic radius ($R$ = Ho, Er, Tm, Yb, Lu, and Y) has a close-packed hexagonal structure with a space group $P6_3cm$ at the low-temperature ferroelectric phase and $P6_3/mmc$ at the high-temperature paraelectric phase [14]. Of further interest, Mn$^{3+}$ ions form a two-dimensional (2D) triangular network, which naturally leads to geometrical magnetic frustration with an antiferromagnetic interaction between nearest-neighbor Mn ions. In addition, the tilting and bucking of the MnO$_5$ bipyramid are reported to be closely related to the ferroelectric transition of $R$MnO$_3$ [15, 16]. It is also a salient feature of the ferroelectric transition that below the ferroelectric transition there is an unmistakable trimerization of Mn atoms indicated by Mn $x$-$position$, not to be confused with the doping concentration of Lu (x) [15]. Interestingly, this trimerization seems to get enhanced below the antiferromagnetic transition through some mechanism. This large distortion of Mn trimers at the antiferromagnetic transition thereby seems to provide a desirable link between magnetism on one hand and ferroelectricity on the other hand, i.e. a magnetoelectric coupling. This observation has since drawn significant attentions and been corroborated by several subsequent experiments [16]. More recently, we also reported an interesting magnon-magnon coupling as well as magnon-phonon one in the spin dynamics for LuMnO$_3$ measured by an inelastic neutron scattering technique [17, 18].

YMnO$_3$ is a typical hexagonal manganite with an antiferromagnetic transition at $T_N$ = 75 K and a ferroelectric transition at $T_C$ = 913 K, [19, 20]. And it shows a considerably large magnetic frustration value of 7.8, which is defined as $f = |\theta_{CW}|/T_N$ [21]: Curie-Weiss temperature $\theta_{CW}$ = -545 K. LuMnO$_3$ is another hexagonal manganite with similar physical properties, except for a different magnetic structure. YMnO$_3$ is known to have a $\Gamma_3$ structure while LuMnO$_3$ has a $\Gamma_4$ structure [22]. As regards the origin of the multiferroic behavior, YMnO$_3$ has been extensively studied by using various techniques, both experimental and theoretical. In particular, high-resolution neutron diffraction studies reported by S. Lee $et$ $al.$ [1] revealed that this system undergoes a giant spin-lattice coupling at the antiferromagnetic transition temperature. There have since been several investigations into this highly unusual behavior and now there is a general agreement about the experimental observation. Yet, the origin and underlying microscopic mechanism of the phenomenon remain still elusive.

Interestingly, while the trimerization of YMnO$_3$ expands below the antiferromagnetic transition temperature: strictly speaking, it is anti-trimerization, but that of LuMnO$_3$ contracts with temperature, i.e. the usual trimerization. One may argue that while the initial formation of the trimerization/anti-trimerization at the ferroelectric transition temperature itself is an intriguing observation it appears to be a far more delicate problem of fundamental importance how this trimerization is coupled to the magnetic order parameter at much lower temperatures. In essence, it is directly related to the question about the exact nature and mechanism of the magnetoelectric coupling of RMnO$_3$. It is also generally believed that the different magnetic structures of YMnO$_3$ and LuMnO$_3$: $\Gamma_3$ and $\Gamma_4$, respectively, are closely related to the different types of the Mn trimerization [1, 22]. These are very surprising results considering a relatively small difference in the ion size of Y and Lu: the ionic radii of Y and Lu are 90 and 86.1 pm, respectively. Experimentally, it implies that the different types of the trimerization induce a different magnetic anisotropy for Mn moments [23].

One of the main surprises in the studies of hexagonal RMnO$_3$, however, is the large spin-lattice coupling found at the antiferromagnetic transition temperature [1]. How this large trimerization/antitrimerization of Mn atoms is related the magnetoelectric coupling and how it affects the underlying physical properties have been the central question of the hexagonal RMnO$_3$. Although the anomaly due to the spin-lattice coupling has been well documented for undoped RMnO$_3$ including both YMnO$_3$ and LuMnO$_3$, it is less well known how this intriguing anomaly varies upon doping. For example, as YMnO$_3$ and LuMnO$_3$ exhibit antitrimerization and trimerziation at $T_N$, respectively, it was speculated [22] that upon doping the anomalous spin-lattice coupling gets weaker and may disappear at a critical doping of Y or Lu. At this critical doping concentration, one might just then be able to realize a perfect triangular lattice without trimerization. This doped system without the Mn trimerization, once synthesized, would make an ideal playground for 2D triangular magnetism. Although the speculation on the doping dependence of the anomaly seems to be natural based on the then known data, this idea has never been put to a rigorous experimental test.

As another way of examining the spin-lattice coupling issue, we can also consider Grüneisen analysis that has been successfully adopted for various problems of similar nature before: one classic example is the issue of quadrupolar ordering and associated strain effects in the context of rare-earth magnetism [24]. To put it simply, Grüneisen parameter captures the response to strain of characteristic temperature and energy scales of a given system. Thus, it has been proven to be a convenient tool for the study of spin-lattice coupling in magnetic materials. Here, we employed the Grüneisen analysis to better understand the spin-lattice coupling issue of hexagonal manganites.

In this paper, we investigated doping effects on the spin-lattice coupling and the trimerization of hexagonal manganite using five single crystals Y$_{1-x}$Lu$_x$MnO$_3$ with x = 0, 0.25, 0.5, 0.75, and 1.0. We are particularly interested in how the lattice and electronic degrees of freedom evolve with doping at the antiferromagnetic transition temperatures. For this, we measured thermal expansion and dielectric constant for both a and c axes.

2. Experiments

We prepared polycrystalline Y$_{1-x}$Lu$_x$MnO$_3$ (x = 0, 0.25, 0.5, 0.75 and 1) using Y$_2$O$_3$, Lu$_2$O$_3$ and Mn$_2$O$_3$ of 99.999 % purity by a standard solid state reaction method. All the starting materials prepared in stoichiometric ratio were mixed and pelletized, and sintered for several times following the recipe as described in Ref. [25]. The final sintering temperature was set at 1300°C for 24 hours. Single crystals were subsequently grown with the feed rod of correct compositions by a floating zone furnace under ambient conditions. The grown single crystal samples were then cut into a cuboid shape with the dimension of 2 mm × 2 mm × 0.5 mm along the c- and a-axes for subsequent bulk measurements. The temperature-dependent x-ray diffraction measurements were also performed from 10 to 300 K using a commercial diffractometer (D8 Advance, Bruker) after grinding the single crystals into powder. We analyzed the diffraction data by using the FullProf program [26] to find that all the samples had no trace of impurity phases and the lattice constants and the unit-cell volume changed linearly following the Vegard's law with increasing Lu concentrations. In order to check the transition temperatures more carefully, we also measured heat capacity and magnetization using two commercial systems (PPMS9 and MPMS5XL, Quantum Design) from 10 to 300 K to find that their respective Neel temperatures are in good agreement with the published results.

We then measured thermal expansion coefficients using a home-made capacitance dilatometer, of which design is similar to that reported in Ref. 27. A major differences between our design and that of Ref. 27 are the facts that our capacitance cell was made of oxygen-free high conductivity (OFHC) copper instead of silver used in Ref. 27 and also was machined to be all-in-one type capacitance cell and electrical shield ring for samples. In our final calibration test, we achieved a resolution of 1.7 Å at 4 K, similar to that reported in Ref. 27. The thermal expansion coefficients measurement was performed by a three-terminal method from 10 to 300 K with a capacitance bridge (2500A, Andeen Hagering) by putting the whole set-up inside a commercial cryostat (PPMS9, Quantum Design). We also measured the dielectric constant from 10 to 300 K with E field along the ab and c axes using a home-made set-up with the same capacitance bridge. For the experiments, we put the samples on top of a gold plate, which was used as an electrode. All the dielectric constants were measured at 1.5 V and 1 kHz.

3. **Results and Discussion**

Figure 1 shows the heat capacity data measured on single crystals $Y_{1-x}Lu_xMnO_3$ (x = 0, 0.25, 0.5, 0.75, and 1). As one can see in Fig. 1(a), the heat capacity data show a clear anomaly at antiferromagnetic transition temperatures consistent with the previous report [22]. For our later discussion on the Grüneisen analysis, we also estimated the magnetic contribution to the heat capacity after subtracting off the phonon contribution using a Debye model with similar Debye temperatures as in Ref. 22 (see inset of Fig. 1(a)). It is noticeable that apart from small variations in the peak positions due to different $T_N$ values there is hardly any change in the shape of the magnetic contribution for all five samples. We note that the total magnetic entropy is about 13 J/mol-$K^2$ for all our sample, close to the theoretical value of 13.38 J/mol-$K^2$. It is also consistent with the fact that our previous neutron diffraction studies found the ordered moment values to remain almost at the same value of 3.3 $\mu_B$/Mn atom across the full doping range [22]. Even the large magnetic heat capacity that persists well into the supposedly paramagnetic phase remains unchanged too. All these indicate that the key features of the two-dimensional triangular magnetism are largely untouched for $Y_{1-x}Lu_xMnO_3$ by doping.

For our later discussion on the Grüneisen analysis, we also measured the temperature dependence of the unit cell volume to find that there is a clear anomaly at $T_N$ as seen for undoped YMnO$_3$ and LuMnO$_3$ (see Fig. 1). What is particularly surprising about our data, however, is that the maximally doped sample with x=0.5 shows almost as clear an anomaly in the temperature dependence of the unit cell volume at $T_N$ as both end compounds although this sample is expected to exhibit a weakest anomaly. This already indicates strongly that we have to revise the earlier idea that somehow there is a crossing point from the anti-trimerization of YMnO$_3$ to the trimerization of LuMnO$_3$ with doping at the rare-earth site.

In order to confirm the anomaly seen for the x=0.5 sample as well as collect the final piece of data for the Grüneisen analysis, we undertook thermal expansion coefficient measurements for both a- and c-axes as shown in Fig. 1(b) and (c). Again, it is very clear that both $\alpha_a(T)$ and $\alpha_c(T)$ display clear anomalies at $T_N$ for all our samples including the x=0.5 sample. Therefore, this observation of the strong anomaly in the data for x=0.5 supports a conclusion that the spin-lattice coupling remains as strong throughout the whole doping range as opposed to the earlier idea. We also note that $\alpha_a(T)$ is positive whereas $\alpha_c(T)$ is negative and the magnitude of $\alpha_a(T)$ is two or three times bigger than the one of $\alpha_c(T)$. That is, the ab plane undergoes far bigger changes due to the magnetic ordering than the c-axis, which is consistent with the ultrasound measurements showing a strong in-plane deformation at $T_N$ [28].

For further exploration of the spin-lattice coupling issue using the thermal expansion data, we decided to estimate the magnetic contribution to the measured total thermal expansion coefficients by subtracting off the phonon contribution. (This analysis provides another useful information on the spin-lattice coupling as we show later.) For the estimate of the phonon contribution, we used a Debye-Grüneisen model: which is a standard model for the thermal expansion due to unharmonic phonon contributions.

According to this model, a typical temperature-dependence of the unit-cell volume can be described [22, 30] by the following formula:

$$V(T) = \frac{V_0 U(T)}{Q - bU(T)} + V_0, \quad (1)$$

where $V_0$ is the unit cell volume at 0 K, $Q = V_0 K_0 \gamma$ and $b = 1/2 (K'_0 - 1)$. $K_0$, $K'_0$ and $\gamma$ are the zero-temperature isothermal bulk modulus, the first derivative of pressure and the thermal Grüneisen parameter, respectively. We used typical values of Q= 3.5 × 10$^{-17}$ J and b= 1.5 throughout our analysis while $V_0$ is a composition-specific value. We note that similar values of Q and b were used for other oxides [29]. The internal energy due to phonon, $U(T)$, is given by the Debye- Grüneisen model [22],

$$U(T) = 9Nk_B T \left(\frac{T}{\Theta_D}\right)^3 \int_0^{\Theta_D/T} \frac{x^3}{e^x - 1} dx, \quad (2)$$

where $\Theta_D$ is the Debye temperature, $n$ is the number of atoms per unit cell and $k_B$ is the Boltzmann constant. The Debye temperatures used in our analysis are 300~500 K, which is similar to those estimated from the analysis of heat capacity [22]. Using this theoretical model, we can calculate the phonon contribution of the temperature-dependent thermal expansion coefficient, $\alpha_V(T)$, by taking the temperature derivative of the temperature dependence of the theoretical volume: $\alpha_V(T) = \frac{1}{V(T)} \frac{dV(T)}{dT}$ (3)

We used the following formula to calculate the total volume thermal expansion coefficients: $\alpha_V(T) = 2 \times \alpha_a(T) + \alpha_c(T)$. In Figs. 2 (a) and (b), we show the measured total volume expansion coefficients for $YMnO_3$ and $LuMnO_3$ together with the pure phonon contributions obtained for both samples following the method as described above. We then took the difference between the two sets of data as the magnetic part of the thermal expansion coefficients, $\alpha_{MAG}(T)$, and show them for all five samples in Fig. 2 (c).

Looking at the thus obtained magnetic part of the thermal expansion coefficients, $\alpha_{MAG}(T)$, it is striking to observe how similar the results look like to the magnetic part of the heat capacity data shown in inset of Fig. 1(a). From the previous work [22], it is known that changes in the unit-cell volume below $T_N$ follow the temperature dependence of the ordered magnetic moment, which was then taken as a strong support for the static spin-lattice coupling in these materials. This new exercise of estimating the magnetic contribution to the thermal expansion coefficients basically confirms the idea. Moreover, the similarity goes well into the supposedly paramagnetic phase, where there are strong dynamic spin fluctuations. This can then be considered as evidence for a dynamic coupling in the paramagnetic phase existing between the lattice and spin degrees of freedom for hexagonal manganites. It is also equally interesting to comment that $Y_{0.5}Lu_{0.5}MnO_3$ shows as strong an anomaly in the magnetic part of the thermal expansion coefficients.

As regards the strong spin-lattice coupling of $(Y,Lu)MnO_3$, it is still illuminating to study the electronic properties such as dielectric constants, not least because of the magnetoeletric coupling in these materials. According to a previous study [13] on undoped $YMnO_3$ and $LuMnO_3$, dielectric constant exhibits a clear anomaly in the in-plane dielectric constant, but not in the c-axis dielectric constant exactly at the antiferromagnetic transition temperatures. This observation seems to be consistent with the strong in-plane deformation reported by the elastic constant measurements [28]. In order to examine how this anomaly in the dielectric constant evolves upon doping, we have measured the dielectric constants of the five single crystals $(Y, Lu)MnO_3$ for both in-plane and c-axis as shown in Fig. 3.

Our new dielectric constants data shown in Fig. 3(a) readily reveal that all our samples show anomalies in the in-plane dielectric constant at $T_N$ with its transition temperatures varying according to doping concentrations. It is further supported by taking the temperature derivative of the dielectric constant as shown in Fig. 3(c). On the other hand, the c-axis dielectric constant does not show any sign of the transition other than a slow decrease across the transition temperature (see the temperature derivative plot in Fig. 3(d)).

One should be reminded that the clear anomalies seen in the in-plane dielectric constants for all the samples including the three doped samples are in good agreement with the observation of similar anomalies in the thermal expansion coefficients as discussed above. However, we should note that our observation seems to sit at variance with one recent dielectric constant measurement, which reported a much weaker change in the dielectric constant for their $Y_{0.7}Lu_{0.3}MnO_3$ sample [31].

The driving mechanism of the magnetoelastic effect of this system seen in both thermal expansion coefficients and dielectric constant was proposed to be due to the displacement of the Mn ions from their ideal symmetric position (x = 1/3), forming a large trimerization distortion [1]. It was also speculated based on then available data that this enhanced trimerization at $T_N$ gets suppressed at the middle of Lu concentration sample like our $Y_{0.5}Lu_{0.5}MnO_3$ [1, 22]. However, this earlier conjecture on possible doping effects on the Mn trimerization and thus the spin-lattice coupling seems to be ill-judged given all our present results: all of them (heat capacity, XRD, coefficient of thermal expansion coefficients, and dielectric constant measurement) show almost as strong spin-lattice coupling effects for all the five (Y, Lu)MnO$_3$ single crystals. Therefore this new observation challenges the earlier view on how the Mn trimerization gets enhanced at the antiferromagnetic transition triggering the magnetoelectric coupling. Although we admit that the actual microscopic mechanism of the trimerization is yet to be understood, however our results provide a clear, strong indication that the Mn trimerization is not a linear function of doping as assumed previously.

We would now like to discuss possible anharmonic effects through the analysis of so-called Grüneisen parameter ($\Gamma$). This Grüneisen parameter ($\Gamma$) is a dimensionless physical quantity that reflects changes in an internal lattice energy with respect to pressure and so strain, $\Gamma$ can be conveniently calculated by using the following formula: $\Gamma = \frac{V_m \cdot K \cdot \alpha_V}{C}$, where $V_m$, $K$, $\alpha_V$, and C are measured molar volume, bulk modulus, volume thermal expansion coefficients, and heat capacity, respectively. In general, $\Gamma$ has the temperature-independent value of 1 – 3 for most common magnetic materials although it can become much larger and/or displays unusual temperature dependence for certain materials [32]. Using the reported bulk modulus value of $K = 1.0 \times 10^{11}$ kg/mT$^2$ for YMnO$_3$ [33, 34], we have calculated the Grüneisen parameter as shown in Fig. 4. Although there is no drastic change at the transition temperature, it is probably safe to say that there is a modest anomaly at $T_N$ for our samples and there is a gradual change as a function of temperature. This continuous change in the Grüneisen parameter even above $T_N$ reflects the fact that there exist both dynamic and static spin-lattice couplings in not only the magnetic phase but also the paramagnetic phase.

Finally, we like to add a general comment on the doping effect of trimerization. The Mn trimerization is one of the major, if not the most important, issues for hexagonal RMnO$_3$. For example, it was pointed out early on that this Mn trimerization triggers the ferroelectricity of hexagonal RMnO$_3$ at very high temperatures via a geometrically frustrated mechanism [35]. This observation directly implies that when the trimerization gets reduced the ferroelectricity itself is also expected to become weaker. The importance of understanding the trimerization issue was further strengthened by several latest very interesting observations of the ferroelectric domain patterns in these materials as well as strong magnon-phonon coupling seen in the spin dynamics data (e.g. see [36, 37]).

In order to help the general public understand better this point, we reproduce some of the data reported in Ref. 1. Figure 5 shows the doping dependence of the Mn *x-position* with respect to the origin of the P6$_3$cm space group at the O3 position. As both end compounds form in the same space group with very similar physical properties, it is natural for one to assume that the Vegard's law will hold here with smooth doping-induced changes in the structural data such as the Mn *x-position*. In that case, there would be a crossover region of doping at around x~ 0.5 – 0.75 as indicated by the thick slanted line in the figure, where the Mn *x-position* crosses the ideal value of 1/3 hence leading to a probable zero trimerization state for that critical doping. The accepted view in the community is that the trimerization is crucial to having a ferroelectric transition in the hexagonal manganites through the so-called geometrically frustrated improper ferroelectricitiy [35]. Therefore, it is expected that a sample without the trimerization would somehow exhibit significant modification its ferroelectric properties and so its magnetoelectric behavior.

However, the key experimental finding reported here is that our measured data behave exactly opposite to this simple view of linear and gradual doping effects. This can be taken as evidence that the trimerization is still as strong for all three doped composition, indicative of some kind of nonlinear doping effects on the trimerization. To be specific, we think that there is some kind of discontinuity in the Mn *x-position* with doping, i.e. no zero crossing. Given the importance of Mn trimerization for two major issues of RMnO$_3$; one is the mechanism of the geometrically frustrated ferroelectricity and another the highly unusual domain physics, our finding of nonlinear doping effect on the trimerization provides a new window of opportunities furthering these two interesting issues using doped systems. Having said that, it may still be just possible although we consider it less likely that despite the Mn *x-position* approaching the critical value of 1/3 much of the physical properties might well remain intact regardless of doping.

## 4. Summary

It has been an intriguing question how the spin-lattice coupling and so the trimerization of the hexagonal manganite evolves upon doping. In order to answer the question directly and understand better the spin-lattice coupling of the hexagonal RMnO$_3$, we synthesized five single crystals Y$_{1-x}$Lu$_x$MnO$_3$ with x = 0, 0.25, 0.5, 0.75, and 1.0. To probe both the lattice and charge degrees of freedom, we measured the thermal expansion coefficients and dielectric constant as a function of temperature. From these measurements, we found that unlike the initial speculations all our samples, including even the maximally doped sample with x=0.5 and 0.75, exhibit clear anomalies at their antiferromagnetic transition temperatures in all the measured data. This observation begs a certain revision to the earlier idea that at an intermediate doping the trimerization gets significantly weakened and they would somehow form a perfect triangular lattice of Mn. Instead, it indicates that the trimerization is as strong for even the doped sample as the undoped pure samples. Our further analysis of Grüneisen parameter shows that there may well be a dynamic spin-lattice coupling in the paramagnetic phase too.

**Acknowledgements**

We acknowledge Martin Rotter for his very helpful discussion on our design and construction of the dilatometer cell used in this study. We would also like to thank Y. M. Song for her technical supports during our thermal expansion measurements at the National Center for Inter-university Research Facility (NCIRF) at Seoul National University. We benefits from useful discussion with Joosung Oh. The work at the IBS CCES was supported by the research program of Institute for Basic Science (IBS-R009-G1).

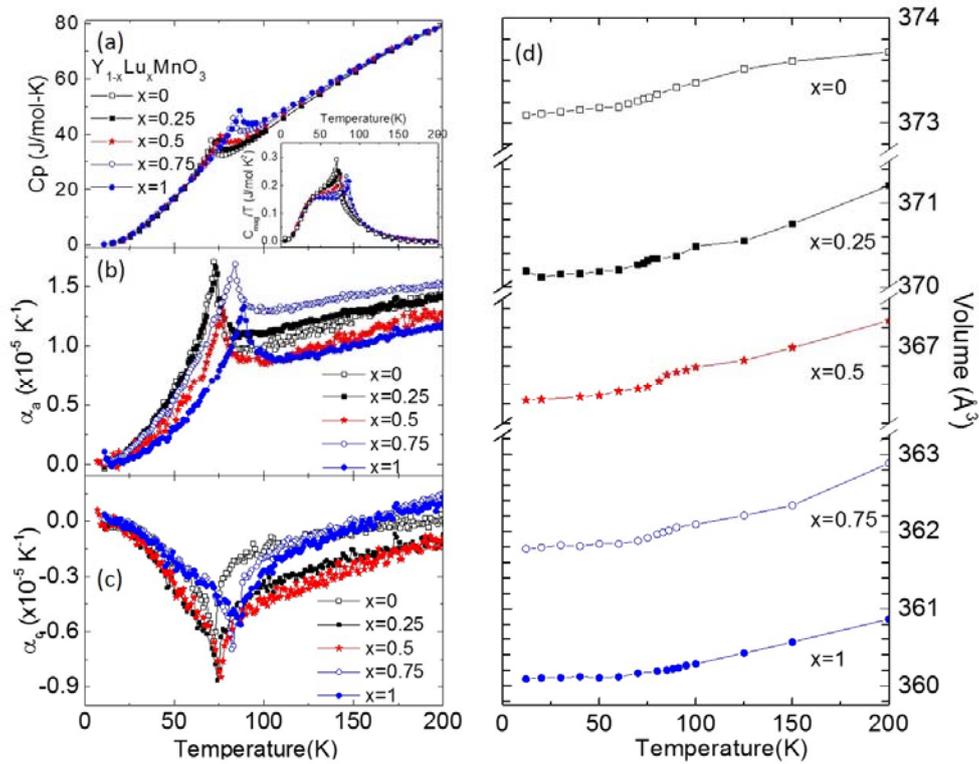

Figure 1. (Color online) Temperature dependence of (a) heat capacity, (b) & (c) thermal expansion coefficients along the a- and c-axes, and (d) the unit cell volume is shown for $Y_{1-x}Lu_xMnO_3$ (x = 0.0, 0.25, 0.5, 0.75, and 1.0). The inset in (a) shows the magnetic contribution of heat capacity for $Y_{1-x}Lu_xMnO_3$.

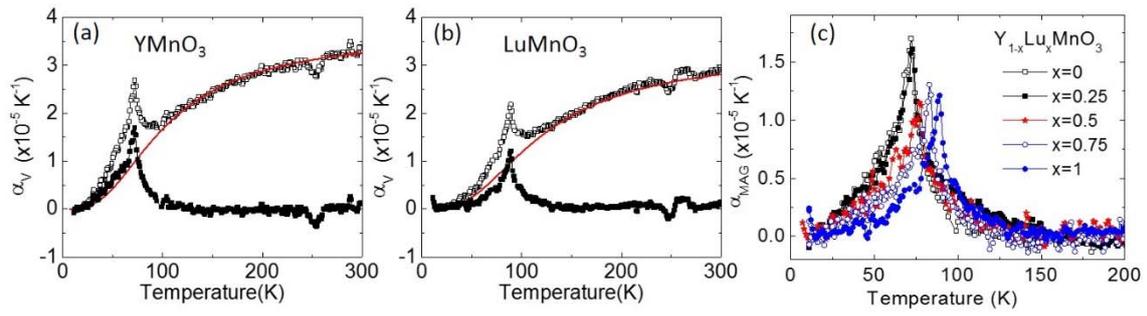

Figure 2. (Color online) It shows the measured (open symbols) thermal expansion coefficients for (a) YMnO$_3$ and (b) LuMnO$_3$ together with our theoretical estimate of phonon contribution (line) using a Debye-Gruneissen model as discussed in the text. The filled symbols represent the difference, which is the obtained magnetic cotribution to the volume thermal expansion coefficients. (c) Temperature dependence of magnetic contributions to the thermal expnaision coefficients is shown for all our samples.

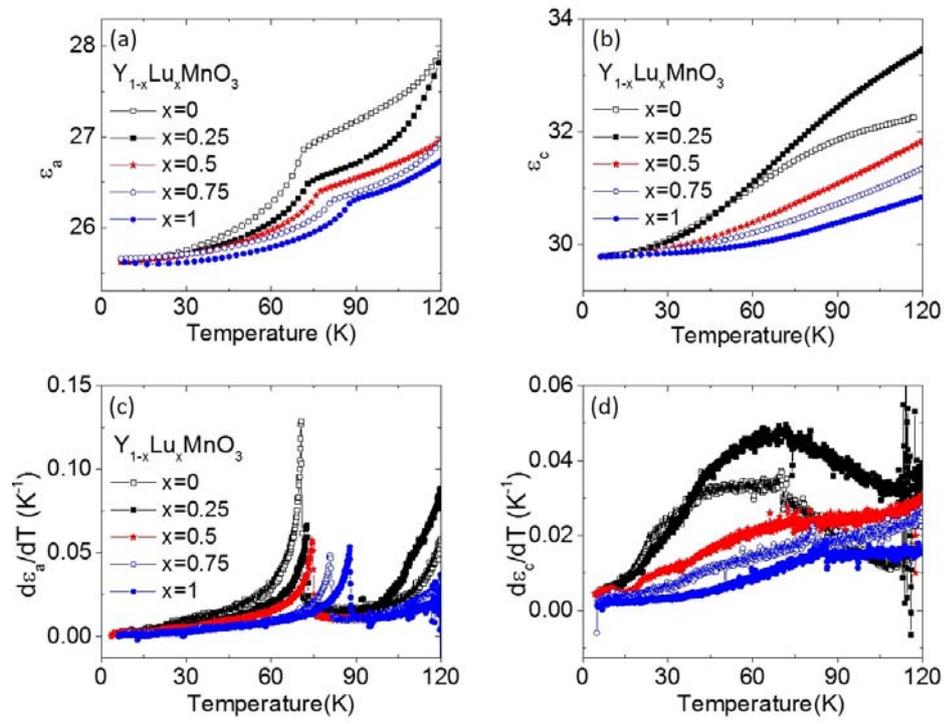

Figure 3. (Color online) Temperature dependence of dielectric constants for (a) a-axis and (b) c-axis together with their temperature derivative shown in (c) and (d).

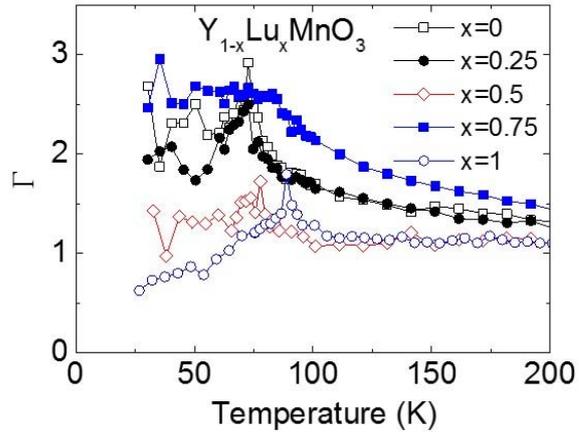

Figure 4. (Color online) Temperature depenence is shown for Grüneisen parameters for all five single crsytals (Y,Lu)MnO$_3$.

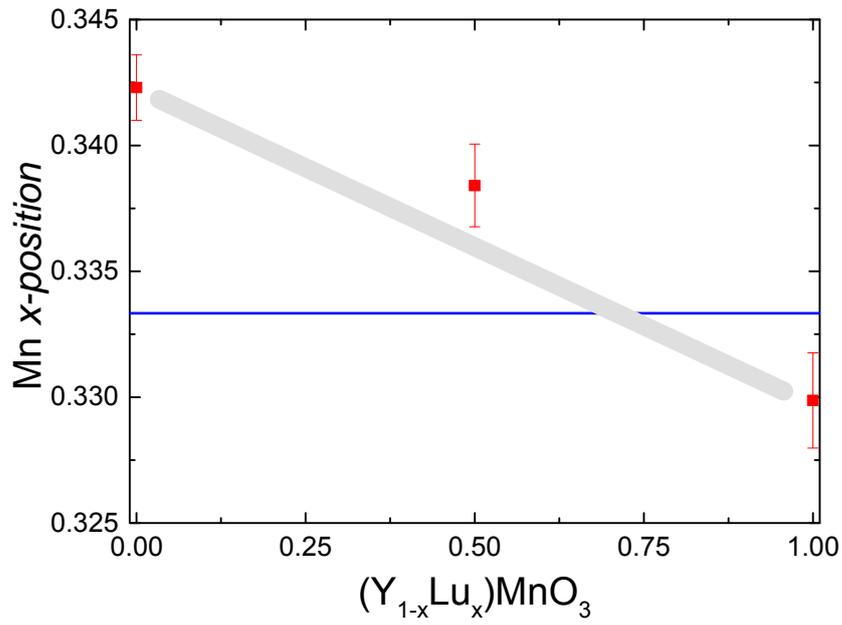

Figure 5. (Color online) Mn *x-position* of $Y_{1-x}Lu_xMnO_3$ system taken after Ref. 1. The symbols represent the position of the Mn *x-position* determined from high-resolution diffraction experiments and the horizontal line indicates the Mn *x-position* (x=1/3) for an ideal triangular lattice of Mn atoms without the trimerization effects. The thick slanted line shows the so-called hypothetical line with a gradual effects of Y/Lu doping with random distribution.